\documentclass[aps,prl,showpacs,superscriptaddress,reprint]{revtex4-1}

\usepackage[T1]{fontenc}

\usepackage{graphicx}
\usepackage{bm,upgreek} 
 
\usepackage{amsmath}
\usepackage{amssymb}

\usepackage{microtype}
\usepackage{hyperref} 

\usepackage{datetime}
\usepackage{color}

\graphicspath{ {./figures/} }

\newcommand{\bvec}{\ensuremath{\bm{\upbeta}}}
\newcommand{\xivec}{\ensuremath{\bm{\xi}}}
\newcommand{\kvec}{\ensuremath{\bm{\upkappa}}}
\newcommand{\mvec}{\ensuremath{\mathbf{m}}}

\newcommand{\ngr}{\ensuremath{\mathcal{N}}}
\newcommand{\avg}[1]{\ensuremath{\langle #1 \rangle}}
\newcommand{\erg}[1]{\ensuremath{\text{ERG}( #1 )}}
\newcommand{\ergm}{\ensuremath{\mbox{ERG}(\mathbf{m})\;}}

\newcommand{\med}{\ensuremath{m_{|}}}
\newcommand{\mts}{\ensuremath{m_{\vee}}}
\newcommand{\mtr }{\ensuremath{m_{\vartriangle}}}

\begin{document}

\title{Reducing Degeneracy in Maximum Entropy Models of Networks}

\author{Szabolcs Horv\'at} 
\affiliation{Department of Physics, University of Notre Dame, Notre Dame, IN, 46556 USA}
\author{\'Eva Czabarka} 
\affiliation{Department of Mathematics, University of South Carolina, Columbia, SC, 29208 USA} 
\author{Zolt\'an Toroczkai}
\affiliation{Department of Physics, University of Notre Dame, Notre Dame, IN, 46556 USA}


\begin{abstract}
Based on Jaynes' maximum entropy principle, exponential random graphs provide a family of principled models that allow the prediction of network properties as constrained by empirical data (observables). However, their use is often hindered by the degeneracy problem characterized by spontaneous symmetry-breaking, where predictions fail. Here we show that degeneracy appears when the corresponding density of states function is not log-concave, which is typically the consequence of nonlinear relationships between the constraining observables. Exploiting these nonlinear relationships here we propose a solution to the degeneracy problem for a large class of systems via transformations that render the density of states function log-concave. The effectiveness of the method is illustrated on examples.
\end{abstract}


\maketitle

Our understanding and modeling of complex systems is always based on partial information, limited data and knowledge. The only principled method of predicting properties of a complex system subject to what is known (data and knowledge) is based on the Maximum Entropy Principle of Jaynes \cite{Jaynes1957,Jaynes1957a}. Using this principle, he re-derived the formalism of statistical mechanics, both classical \cite{Jaynes1957} and the time-dependent quantum density-matrix  formalism \cite{Jaynes1957a}, using Shannon's information entropy \cite{Shannon1948}.  The method generates a probability distribution $P(\mu)$ over all the possible (micro)states $\mu$ of the system by maximizing the entropy $S[P] = - \sum_{\mu} P(\mu) \ln P(\mu)$ subject to what is known, the latter expressed as ensemble averages over $P(\mu)$.  In this context the given data and the available knowledge act as \emph{constraints}, restricting the set of candidate states describing the system.  $P(\mu)$ is then used via the usual partition function formalism  to make unbiased predictions about other observables. 

The applicability of Jaynes's method extends well beyond physics \cite{Presse2013}, and in particular, it has been applied in biology \cite{Livesey1987,Steinbach2002,Yeo2004,Watkins2006,Vergassola2007,Saul2007,Walczak2010,Morcos2014}, neuroscience \cite{Shlens2006, Schneidman2006,Tang2008,Marre2009,Yeh2010, Stephens2010,Ganmor2011,Watanabe2013,Ercsey-Ravasz2013}, ecology \cite{Phillips2006, Harte2011}, sociology \cite{Fronczak2007,Wimmer2010}, economics \cite{Bass1974,Agrawal2008raey},  engineering \cite{Gull1978, Skoglund1996}, computer science \cite{Rosenfeld1996}, etc. It also received attention within network science  \cite{JAmStatAssoc_HollandL1981,Strauss1986,Park2004a,Park2004,Park2005,Robins2007,Fronczak2013,House2014}, leading to a class of models known as exponential random graphs (ERG). Despite its popularity, however, this method often presents a fundamental problem, the \emph{degeneracy problem}, that seriously hinders its applicability \cite{Park2004,Park2005}.  When this problem occurs,  $P(\mu)$ lacks concentration around the averages of the constrained quantities and the typical microstates do not obey the constraints. In case of ERGs, the generated graphs, for example, may either be very sparse, or very dense, but hardly any will have a density close to that of the data network. Predictions based on such distributions can be significantly off. Two basic questions arise related to the degeneracy problem: 1) Under what conditions it occurs? and 2) How can we eliminate or minimize this problem?

In this Letter we answer both questions and present a solution that significantly reduces degeneracy, then illustrate its effectiveness on concrete examples. We will present our analysis and results using the language of networks and ERG models, however, our findings are generally applicable. Let us consider the set ${\cal G}_N$ of all labeled simple graphs $G \in {\cal G}_N$ (no parallel edges, or self-loops) on $N$ nodes, corresponding here to microstates $\mu$, and an arbitrary set of graph measures or observables $\mvec(G) = m_1(G),\ldots,m_K(G)$, e.g., the number of edges $m_|$, 2-stars $\mts$, triangles $\mtr$, the degree of the 9th node. These measures represent the constraints and we assume that we are given specific values $\mvec^0$, for them (input data). They may come from an empirical network $G^0$, or could represent averages from several empirical datasets.  A key assumption in Jaynes' method is to impose these data at the level of ensemble averages:
\begin{equation}
\mvec^0 = \langle \mvec(G) \rangle = \sum_{G \in {\cal G}_N} 
\mvec(G) P(G)\;, \label{ensav}
\end{equation}
and the goal is to determine the ensemble itself, i.e., the probabilities $P(G)$ for all $G$, as constrained by (\ref{ensav}) and normalization: $\sum_{G \in {\cal G}_N} P(G) = 1$.  Since the number of constraints $K$ is usually small, system (\ref{ensav}) is strongly underdetermined, the number of unknowns being $| {\cal G}_N | = 2^{{\cal O}(N^2)}$.  Following Jaynes, the least biased distribution $P(G)$ obeying the constraints is the one that maximizes the entropy $S[P] = - \sum_{G \in {\cal G}_N} P(G) \ln P(G)$ subject to (\ref{ensav}) and normalization. The method of Lagrange multipliers then yields the family of Gibbs distributions:
\begin{equation}
P(G)=P(G;\bvec) = 
\frac{e^{- \sum_{k=1}^K \beta_k \, m_k(G)}}{Z(\bvec)} =
\frac{e^{-\bvec\cdot \mvec(G)} }{Z(\bvec)}\;, \label{gibbs}
\end{equation}
where $Z(\bvec) = \sum_{G \in {\cal G}_N} e^{-\bvec\cdot \mvec(G)}$ is the partition function. The $\bvec = (\beta_1,\ldots,\beta_K)$ are Lagrange multipliers associated with the constraints $\mvec = (m_1,\ldots,m_K)$,  determined from solving system (\ref{ensav}) with (\ref{gibbs}), i.e.,  
\begin{equation}
\langle m_k \rangle = \frac{\partial F(\bvec)}{\partial \beta_k} \label{betam}
\end{equation}
where $F(\bvec) = -\ln Z(\bvec)$ denotes the free energy. 
The average of some other graph measure $q(G)$ in this ensemble will be $\langle q \rangle = \sum_{G \in {\cal G}_N} q(G) P(G;\bvec)$. The distribution $P(G;\bvec)$ defines the corresponding exponential random graph model, hereinafter referred to as the $\mbox{ERG}(\mvec)$ model.  Eq. (\ref{betam}) admits a maximum likelihood interpretation: its solution is the set of parameters \bvec\ that maximize the probability $P(G^0;\bvec) = Z^{-1}(\bvec)e^{-\bvec\cdot \mvec^0}$ of the graph $G^0$ for which $\mvec(G^0) = \mvec^0$. Note that all graphs  having the same properties $\mvec$ will have the same probability in the \erg{\mvec} model.

Since the partition function is determined by the graph measures only, we may write
$Z(\bvec) =  \sum_{\mvec} {\cal N}(\mvec)e^{-\bvec\cdot \mvec}$,
where  ${\cal N}(\mvec)$ is a \emph{counting function}, representing the number of graphs that have the same values for these measures, equivalent to the density of states function in physics. For example,  ${\cal N}(m_{\mid},\mtr)$ is the number of graphs with $m_{\mid}$ edges and $\mtr$ triangles.  
To simplify the notations, in the following we will work with adimensional and rescaled quantities $m_i \in [0,1]$ \footnote{In the figures, however, we indicate the full range of the values.}. Let us denote the domain of ${\cal N}$ by ${\cal D} = \{\mvec \in [0,1]^K \, | \, {\cal N}(\mvec) \geq 1\}$.  Therefore, the probability that a graph sampled by the $\mbox{ERG}(\mvec)$ model will have the given $\mvec$ is: 
\begin{equation}
p(\mvec;\bvec) = \frac{{\cal N}(\mvec)}{Z(\bvec)} e^{-\bvec \cdot \mvec}\;,  
\label{pm}
\end{equation}
and thus we can write (\ref{betam}) as the mean of $p(\mvec;\bvec)$:
\begin{equation}
\langle \mvec \rangle = \sum_{\mvec} \mvec\,p(\mvec;\bvec)\;. \label{mkp}
\end{equation}

\textit{Sharp constraints.}---In the above the constraints were imposed at the level of averages. It may happen, however,  that some of the data holds for all states of the system, akin to integrals of motion in physics. In network science in this case we restrict ourselves to the largest set of graphs ${\cal G}_N(\mvec^0) \subseteq {\cal G}_N$, all having the same value $\mvec^0$ for those particular measures. We refer to these types of constraints as \emph{sharp constraints}. Examples include the set of all graphs with a given number of edges (the  $G(N,M)$ model), introduced by Erd\H{o}s and R\'enyi \cite{RndGrphs_Bollobas2001},  or those with a given degree sequence \cite{JPhysA_KimTMES2009, PLoSONE_DelGenioKTB2010}, or with given joint-degree matrix \cite{ArXiv_CzabarkaDEM2013}. While sharp constraint problems are mathematically hard in general, counting problems, i.e., computing ${\cal N}(\mvec)$, were shown to be the hardest \cite{TheorCompSci_JerrumVV1986,  ApproxAlg2003}. 

\begin{figure}[htbp]
  \begin{center}
    \includegraphics{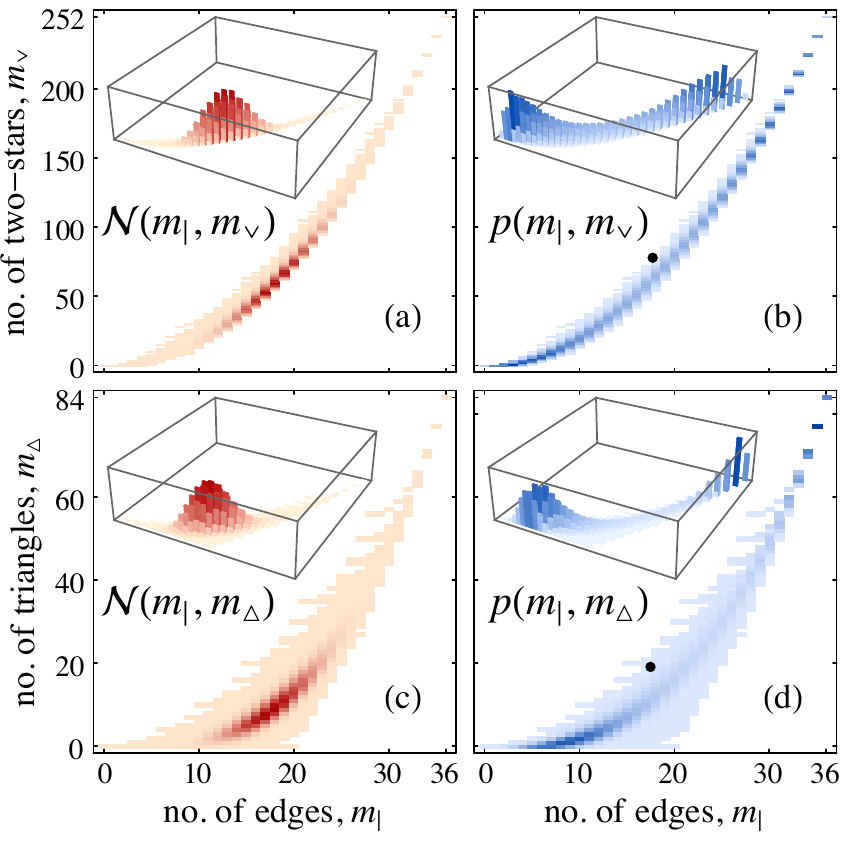}
    \vspace*{-0.5cm}
    \caption{Degenerate ERG models. Plots are from exact enumeration of all labeled graphs on $N=9$ nodes.  (a)  The counting function $\ngr(\med,\mts)$. Color intensity is proportional to the value of $\ngr$, white means ${\cal N} = 0$ there.  (b) Distribution  $p(\mvec;\bvec)$ from \erg{\med,\mts} at $\beta_{|}^0=2.20$ and $\beta_\vee^0=-0.313$, corresponding to the black dot. (c) $\ngr(\med,\mtr)$. (d) $p(\mvec;\bvec)$ from \erg{\med,\mtr} with $\beta_{|}^0=1.24$ and $\beta_\vartriangle^0=-0.610$ from the black dot. Insets show 3D versions of the intensity plots. Note from (\ref{pm}) that the domains of ${\cal N}$ and $p$ always coincide.}
    \label{fig1}
    \vspace*{-0.5cm}
  \end{center} 
\end{figure}

\textit{The degeneracy problem.}---When solving (\ref{betam}) (or (\ref{mkp})) for $\bvec$ with given $\langle \mvec \rangle = \mvec^0$ we are fixing the parameters $\bvec(\mvec^0) \equiv \bvec^0$. It may happen that $p(\mvec;\bvec^0)$ is multimodal, with probability mass concentrated around two or more disjoint and well separated (by ${\cal O}(1)$ distances) domains in the observables $\mvec$, in which case the  \ergm  is called degenerate.  As examples, let us consider the two ERG models, \erg{\med,\mts} and \erg{\med,\mtr}, shown in Fig.~\ref{fig1}.  Figures~\ref{fig1}(b),~\ref{fig1}(d) show  $p(\mvec;\bvec)$ at parameter values corresponding to averages $(\langle m_1 \rangle, \langle m_2 \rangle)$ indicated by the black dots. We see that both models are degenerate: for these input values (or corresponding parameters), the sampled graphs will be either very dense or very sparse, practically none with observable values similar to the input data. 
This is true even in the case when the averages are realizable by specific graphs (seen more clearly in Fig.~\ref{fig1}(d)).  Observe that the $\langle \mvec \rangle$ averages can come from any point in the convex hull of ${\cal D}$ (and only from there). Also note that in both cases ${\cal N}(\mvec)$ itself is unimodal, however, $p(\mvec; \bvec^0)$ is multimodal
\footnote{See Supplemental Material at [URL will be inserted by publisher], which includes Refs.~\cite{Ramond2010,Bona2004,Agrawal2008raey,Gleiser2003}}.
\nocite{Ramond2010,Bona2004,Agrawal2008raey,Gleiser2003}
It is important to emphasize that when degeneracy occurs the graphs sampled by $p(\mvec;\bvec)$ are coming from regions with significant probability mass whose separation is large, \emph{comparable to unity}.  Strictly speaking, ${\cal N}(\mvec)$ is a combinatorial function and it may be jagged locally (integer effects).  However, samples from nearby peaks are similar, which is fine for modeling purposes, it is not considered degenerate. For that  reason, (keeping the notation) in the remainder we will refer to the \emph{smoothened, continuous} version of ${\cal N}(\mvec)$, preserving only its long-wavelength properties. 
For another, non-network example of a degenerate maximum entropy model see \cite{Note2}.
Degeneracy can be best understood in 1D, $K = 1$.  Let $f : [a, b] \rightarrow \mathbb{R}^+$ be a twice differentiable positive function, and let $g(x) = f(x) e^{- \beta x}$. Since $g(x) > 0$, the condition for $g(x)$ not to be multimodal for any $\beta$ is that it should not have any minima in $(a,b)$ for any $\beta$. This is true if in any stationary point $x_0$, i.e., with $g' (x_0) = 0$, the function $g$ is concave, $g''(x_0) < 0$. For a stationary point $x_0$ we have $\beta = f' (x_0) / f(x_0)$. Computing $g''(x_0)$ and eliminating $\beta$ from it using the above, we get $f''(x_0) f(x_0) < f' (x_0)^2$. Any $x_0 \in (a,b)$ can be stationary, since $f(x_0) > 0$ and thus the corresponding $\beta = f' (x_0) / f(x_0)$ always exists to make $x_0$ stationary. Therefore, $g(x)$ will be non-degenerate if and only if $f''(x) f(x) - f'(x)^2 <  0$ for all $x\in(a,b)$.  This is, however, equivalent to saying that $f(x)$ is strictly \emph{log-concave}, i.e., $\ln f(x)$ is (strictly) concave: $d^2 (\ln f(x))/dx^2  < 0$ for any $x \in (a, b)$.  For example, Gaussians are log-concave. 
Generalizing this for arbitrary dimensions (for proof see \cite{Note2}), we can announce:

\noindent \emph{\underline{Theorem}:} The \erg{\mvec}  is non-degenerate if and only if the 
density of states ${\cal N}(\mvec)$ is strictly log-concave.

The necessary and sufficient conditions for function ${\cal N}(\mvec)$ to be log-concave \cite{ConvexOpt_BoydV2004} is that (\textit{i}) its domain ${\cal D}$ is convex and (\textit{ii}) if (\textit{i}) holds, to satisfy the Pr\'ekopa--Leindler type inequality ${\cal N}(\lambda \mvec + (1-\lambda) \mathbf{n}) > {\cal N}(\mvec)^{\lambda}  {\cal N}(\mathbf{n})^{1-\lambda}$ for any $1 < \lambda < 1$ and $\mvec,\mathbf{n} \in {\cal D}$ 
\footnote{Equivalent to the usual definition for strict concavity $g(\lambda \mathbf{x}_1 + (1-\lambda) \mathbf{x}_2) < t g(\mathbf{x}_1) + (1-t)g(\mathbf{x}_1)$, $\forall \mathbf{x}_i \in {\cal D}, \lambda \in (0,1)$ with $g = \ln{\cal N}$}. 
It is important to note that the theorem above reduces degeneracy to purely graph theoretical properties.
In two or higher dimensions degeneracy occurs frequently, and the typical approach has been simply to switch to an entirely different set of measures \cite{Snijders2006}.  Realistically, however, we might not have other data, or its collection would not be an option; we want to extract the maximum possible information from the available data. Additionally, from a domain expertise point of view, e.g.,  triangle count is a natural variable for sociologists, as it expresses the level of transitivity, an important measure for social networks;  yet the corresponding  ERG model is degenerate \cite{Strauss1986}.

\begin{figure}[htbp] 
  \begin{center}
    \includegraphics{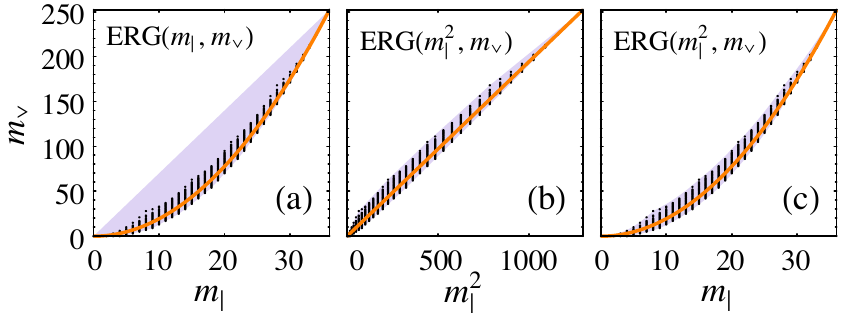} 
    \vspace*{-0.7cm}
    \caption{Domain ${\cal D}$ (black dots) and its convex hull (purple shading); the averages \avg{\mvec} take their values from the convex hull. The orange line is $\mts \sim \med^2$. (a) \erg{\med,\mts}, in $(\med,\mts)$ space. Note the large region of possible \avg{\mvec} values with no realizable graphs (no black dots).  (b) \erg{\med^2,\mts}, in $(\med^2,\mts)$ space. Now ${\cal D}$ and its convex hull almost coincide.  (c) \erg{\med^2,\mts} in $(\med,\mts)$, compare with (a). }
    \label{fig2} 
    \vspace*{-0.5cm} 
  \end{center} 
\end{figure}

\begin{figure}[htbp] 
  \begin{center}
    \includegraphics{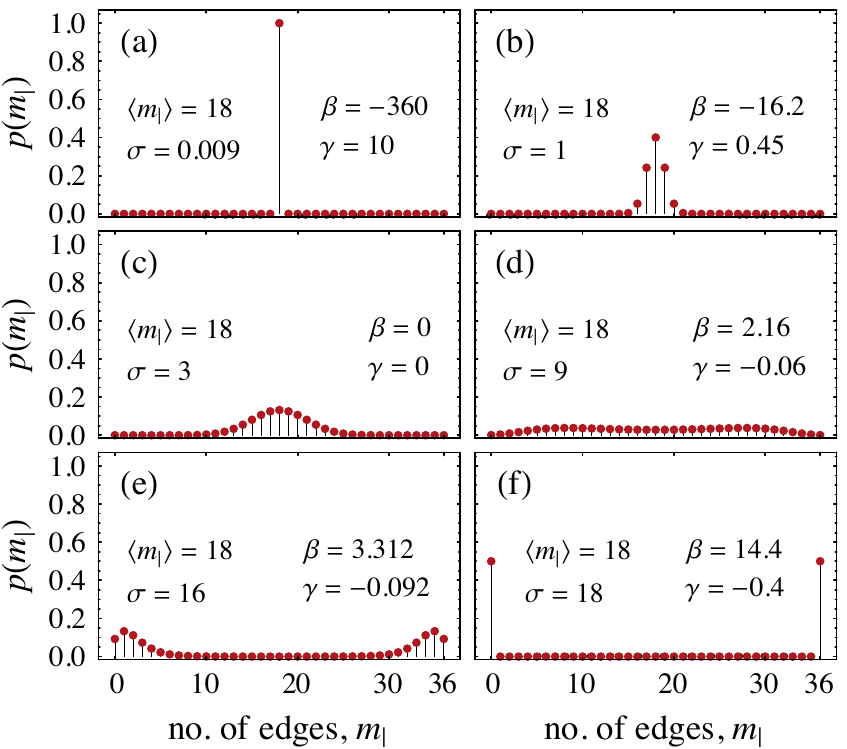}
    \vspace*{-0.4cm}
    \caption{Distribution of the edge count $m_|$ of the sampled graphs ($N=9$ nodes) in the \erg{\med, \med^2} model at various parameter values, where $p(\med) \propto \ngr(\med) \exp{(-\beta \med - \gamma \med^2)}$.}
    \label{fig3} 
    \vspace*{-0.5cm} 
  \end{center} 
\end{figure}

\begin{figure}[htbp]
  \begin{center}
    \vspace{-0.3cm}
    \includegraphics{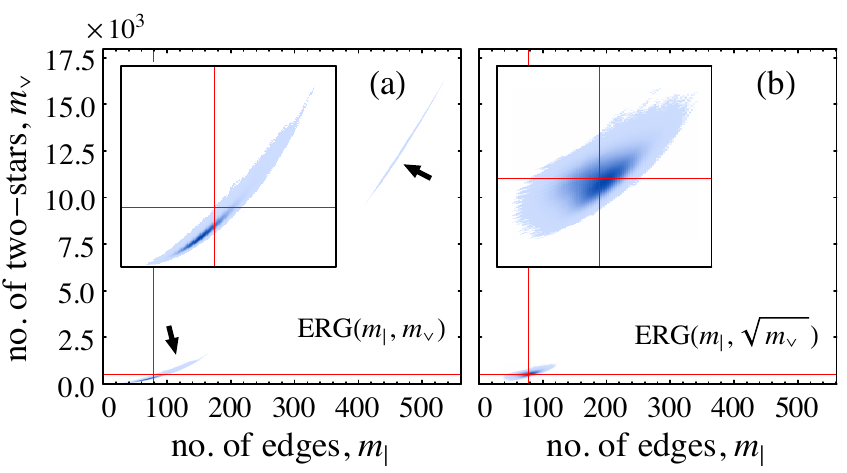}
    \includegraphics{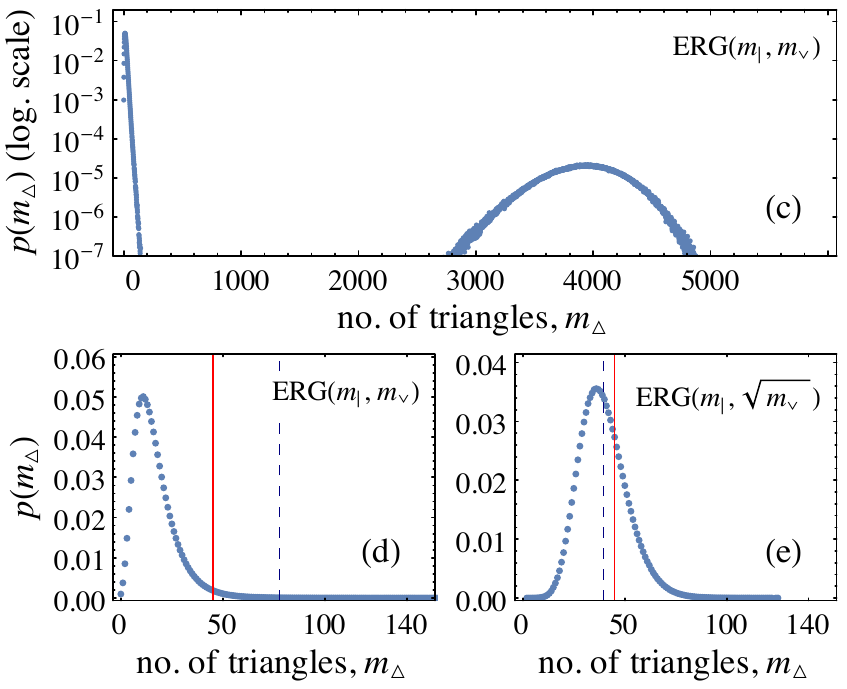}
    \vspace{-0.3cm}
    \caption{Modeling the Zachary Karate Club data. Distributions for \erg{\med, \mts} ((a), (c), (d)) and \erg{\med, \sqrt{\mts}} ((b), (e)) after fitting.  (a) $p(\med,\mts)$ in \erg{\med,\mts} and (b) $p(\med,\mts)$ in \erg{\med, \sqrt{\mts}}.  The cross-hair is at $(\med^0, \mts^0)$. Insets are magnifications around $(\med^0, \mts^0)$. Arrows (a) indicate the two modes of the degenerate distribution. (c)-(e) show $p(\mtr)$ in the two models. The red vertical lines are at $\mtr^0$ and the dashed ones are model averages.}
    \vspace{-0.6cm}
  \label{fig4}
  \end{center}
\end{figure}

\textit{Solution.}---Here we propose to work still with the same variables $\mvec$ (same data) as in the degenerate ERG model, however, to consider a \emph{one-to-one}  transformation $\mvec \leftrightarrow \xivec = \mathbf{F}(\mvec)$ such that the corresponding counting function:
\begin{equation}
\overline{{\cal N}}(\xivec) = {\cal N}(\mathbf{F}^{-1}(\xivec))
\end{equation}
is log-concave \footnote{Since $\mathbf{F}$ is bijective, the counts of states with $\bm\xi$ is the same as the the counts of states with $\mvec=\mathbf{F^{-1}}(\bm\xi)$.}. Due to the one-to-one nature, one can still work with or plot the distributions in the same coordinate system $\mvec$ (see Fig.~\ref{fig2}(b)(c)), but the graphs are sampled by the non-degenerate model \erg{\xivec} = \erg{\mathbf{F}(\mvec)},  with constraints $\xivec^0  = \mathbf{F}(\mvec^0) = \langle \xivec \rangle$.  There is no recipe for obtaining such transformation in general (it might even not exist, e.g., when ${\cal D}$ is not singly connected), however, there is a large class of problems where this can be achieved, to which the degenerate models in the literature belong.
This is the case when the convexity condition (\textit{i}) is violated.  To better understand the nature of the $\mathbf{F}$ function in this situation, let us focus on the 2D case.  If $m_1(G)$ and $m_2(G)$ were independent,  ${\cal D}$ would be rectangular and therefore convex.  Instead, the shapes of the domains in Fig.~\ref{fig1} indicate that there is a nonlinear confining relationship between the variables, on average.
For the $(m_|,\mts)$ case it holds that $\mts \sim m_|^2$ on average (Fig.~\ref{fig2}(a), thick orange line).  Similarly, for $(m_|,\mtr)$ we have $\mtr \sim m_|^3$ (not shown).  Focusing on the $(m_|,\mts)$ case we can pinpoint why such nonlinear dependencies cause degeneracy. Since $\mts \sim m_|^2$, choosing the constraints arbitrarily we are \emph{independently} setting both the average of $m_|$ and its spread $\sigma = (\langle m_|^2 \rangle - \langle m_| \rangle^2)^{\frac{1}{2}}$. 
This is shown most directly by looking at an \erg{m_|,m_|^2} model (see Fig.~\ref{fig3}). Since the network is finite, the spread $\sigma$ can be tuned from a small value corresponding to a unimodal distribution for $m_|$, Fig.~\ref{fig3}(a)-\ref{fig3}(c), to its maximum Fig.~\ref{fig3}(d)-\ref{fig3}(f), where the probability mass is bimodal, hence causing degeneracy. Note, a linear relation between the variables will not cause degeneracy.
 This suggests to choose $\mathbf{F}$ such as to convexify the domain via linearization, i.e., to have $\xi_1 \sim \xi_{2}$. For example, for the $(m_|,\mts)$ case this could be done via  $\xi_| = m_|^{2\theta}$, $\xi_{\vee} = \mts^{\theta}$, with $\theta > 0$ arbitrary, as shown in Fig.~\ref{fig2}(b) for $\theta = 1$, or for $\theta = 1/2$ in the model of Fig.~\ref{fig4}. 

Recall that in the original (degenerate) \erg{\mvec} we had $\langle \mvec \rangle = \mvec^0$ precisely, by definition. However, the new model \erg{\xivec} is constrained by $\langle \xivec \rangle_{\xi} = \mathbf{F}(\mvec^0) \equiv \xivec^0$, where the subscript $\xi$ indicates averages in \erg{\xivec}. Here $\langle \mvec \rangle_{\xi} \neq \mvec^0$, yet $\langle \mvec \rangle_{\xi} \approx \mvec^0$ will hold. Let $\kvec^0$ denote the Lagrange parameters in the \erg{\xivec} model. For the $i$th component, the difference is on the order of $\frac{1}{2} |\sum_{\xivec}(\xivec - \xivec^0)^T H[F_i^{-1}](\xivec^0)(\xivec - \xivec^0) p(\xivec;\kvec^0)|  \leq \frac{K}{2} \| H[F^{-1}_i](\xivec^0)\|_2 \; \|\operatorname{Cov}(\xivec,\xivec)\|_2$, where $H[F^{-1}_i](\xivec^0)$ is the Hessian of $F^{-1}_i(\xivec)$ computed in $\xivec^0$ and $\|\cdot\|_2$ is the spectral norm. Since \erg{\xivec} is non-degenerate, $p(\xivec;\kvec^0)$ will be concentrated around $\xivec^0$, in a region small compared to unity, and additionally, over this region the variability of $\mathbf{F}$ is small ($\mathbf{F}$ straightens the whole domain ${\cal D}$, varying significantly only over ${\cal O}(1)$ distances). Thus, while this transformation leads to minor differences, it resolves the degeneracy problem and the samples are with high probability from the neighborhood of graphs for which the given constraints are typical.  

\textit{Validation.}---In the following we test the method on Zachary's well-known karate club (ZKC) dataset \cite{Zachary1977}, which describes a network $G^0$ of club friendships (\cite{Note2} shows the test for another network \cite{Gleiser2003}). ZKC has $N = 34$, $\med^0=78$, $\mts^0=528$ and $\mtr^0 =45$. Using Markov Chain Monte Carlo (MCMC) sampling and a stochastic root finding method, we fitted the \erg{m_|,\mts} model to $G^0$ obtaining $\beta_|^0=2.610$, $\beta_{\vee}^0 = -0.08125$ and a degenerate $p(\med,\mts;\beta_|^0, \beta_{\vee}^0)$, shown in Fig.~\ref{fig4}(a).   

Next we fitted the model \erg{\xi_{|} = m_|, \xi_{\vee} = \sqrt{\mts}}, obtaining $ \kappa_{|}^0 = 3.625$ and $\kappa_{\vee}^0 = -7.998$ and a non-degenerate distribution $p(\med,\mts; \kappa_{|}^0, \kappa_{\vee}^0)$, shown in Fig.~\ref{fig4}(b).  The averages are summarized in Table~\ref{tab:karate}.  Even though here we solve for $\langle \sqrt{\mts} \rangle_{\xi} = \sqrt{\mts^0}$, we expect that $\langle \mts \rangle_{\xi} \approx \mts^0$.  This is confirmed in the $\langle \mts \rangle$ column of Table~\ref{tab:karate}. Note that due to the degeneracy of \erg{m_|,\mts}, its prediction for $\langle \sqrt{\mts} \rangle^{2}$ is $370$, far from $528$, whereas  \erg{m_|,\sqrt{\mts}} predicts all quantities well. 

Let us now consider the number of triangles \mtr.  To the extent in which $\med^0$ and $\mts^0$  determine  \mtr, the corresponding ERG model should predict \mtr  as well.  Unsurprisingly, \erg{\med,\mts} produces a bimodal distribution $p(\mtr)$, Fig.~\ref{fig4}(c)-(d) and predicts $\langle \mtr \rangle = 78$, far from 45. Additionally, 45 and 78 are produced with low probability in the \erg{\med,\mts} model (see  Fig.~\ref{fig4}(d)). The \erg{m_|,\sqrt{\mts}} convexified model, however, predicts $\langle \mtr \rangle_{\xi} = 40$, and both 40 and 45 are produced with high probability in this model, see Fig.~\ref{fig4}(e).
\begin{table}[htbp]
  \begin{center} 
    \vspace*{-0.4cm}
    \begin{tabular}{l|c|c|c|c} 
    & \avg{\med} & \avg{\mts} & $\avg{\sqrt{\mts}}^2$ & \avg{\mtr } \\
    \hline
    $G^0$ (ZKC) & 78 & 528 & 528 & 45 \\
    ERG($\med,\mts$) & $77.8 \pm 0.5$ & $530 \pm 9$ & $370 \pm 4$ & $77.7 \pm 2.3$ \\
    ERG($\med,\sqrt{\mts}$) & $77.9 \pm 0.5$ & $530.7 \pm 2.7$ & $527.3 \pm 2.5$ & $39.5 \pm 0.3$ \\
    \end{tabular}
  \end{center} 
  \vspace*{-0.5cm}
  \caption{Averages of measures in the fitted ERG models. $G^0$ denotes the ZKC network. For the averages we also indicate the standard error of the MCMC estimates.}
  \label{tab:karate}
\end{table}

It is important to note that the degeneracy problem, the reason for its occurrence, and the solution proposed here are general, applicable beyond network modeling. We have shown that degeneracy will typically appear when the constraining observables (input data) are nonlinearly constraining one
another so that the density of states function is not log-concave. To avoid degeneracy, but still be able to use the same input data, here we proposed one-to-one mappings of the observables (so that no information
is lost) in ways that render the density of states function log-concave. 


\begin{acknowledgments}
We thank L.~Sz\'ekely, K.~Bassler, M.~Varga and D.C. Vural for discussions.  This work was supported in part by grant No.~FA9550-12-1-0405 of the U.S.\ Air Force Office of Scientific Research, the Defense Advanced Research Projects Agency and the Defense Threat Reduction Agency Award HDTRA 1-09-1-0039. 
\end{acknowledgments}


\bibliographystyle{apsrev4-1}
\bibliography{erg}

\begin{thebibliography}{54}%
\makeatletter
\providecommand \@ifxundefined [1]{%
 \@ifx{#1\undefined}
}%
\providecommand \@ifnum [1]{%
 \ifnum #1\expandafter \@firstoftwo
 \else \expandafter \@secondoftwo
 \fi
}%
\providecommand \@ifx [1]{%
 \ifx #1\expandafter \@firstoftwo
 \else \expandafter \@secondoftwo
 \fi
}%
\providecommand \natexlab [1]{#1}%
\providecommand \enquote  [1]{``#1''}%
\providecommand \bibnamefont  [1]{#1}%
\providecommand \bibfnamefont [1]{#1}%
\providecommand \citenamefont [1]{#1}%
\providecommand \href@noop [0]{\@secondoftwo}%
\providecommand \href [0]{\begingroup \@sanitize@url \@href}%
\providecommand \@href[1]{\@@startlink{#1}\@@href}%
\providecommand \@@href[1]{\endgroup#1\@@endlink}%
\providecommand \@sanitize@url [0]{\catcode `\\12\catcode `\$12\catcode
  `\&12\catcode `\#12\catcode `\^12\catcode `\_12\catcode `\%12\relax}%
\providecommand \@@startlink[1]{}%
\providecommand \@@endlink[0]{}%
\providecommand \url  [0]{\begingroup\@sanitize@url \@url }%
\providecommand \@url [1]{\endgroup\@href {#1}{\urlprefix }}%
\providecommand \urlprefix  [0]{URL }%
\providecommand \Eprint [0]{\href }%
\providecommand \doibase [0]{http://dx.doi.org/}%
\providecommand \selectlanguage [0]{\@gobble}%
\providecommand \bibinfo  [0]{\@secondoftwo}%
\providecommand \bibfield  [0]{\@secondoftwo}%
\providecommand \translation [1]{[#1]}%
\providecommand \BibitemOpen [0]{}%
\providecommand \bibitemStop [0]{}%
\providecommand \bibitemNoStop [0]{.\EOS\space}%
\providecommand \EOS [0]{\spacefactor3000\relax}%
\providecommand \BibitemShut  [1]{\csname bibitem#1\endcsname}%
\let\auto@bib@innerbib\@empty
\bibitem [{\citenamefont {Jaynes}(1957{\natexlab{a}})}]{Jaynes1957}%
  \BibitemOpen
  \bibfield  {author} {\bibinfo {author} {\bibfnamefont {E.~T.}\ \bibnamefont
  {Jaynes}},\ }\href {\doibase 10.1103/PhysRev.106.620} {\bibfield  {journal}
  {\bibinfo  {journal} {Phys. Rev.}\ }\textbf {\bibinfo {volume} {106}},\
  \bibinfo {pages} {620} (\bibinfo {year} {1957}{\natexlab{a}})}\BibitemShut
  {NoStop}%
\bibitem [{\citenamefont {Jaynes}(1957{\natexlab{b}})}]{Jaynes1957a}%
  \BibitemOpen
  \bibfield  {author} {\bibinfo {author} {\bibfnamefont {E.~T.}\ \bibnamefont
  {Jaynes}},\ }\href {\doibase 10.1103/PhysRev.108.171} {\bibfield  {journal}
  {\bibinfo  {journal} {Phys. Rev.}\ }\textbf {\bibinfo {volume} {108}},\
  \bibinfo {pages} {171} (\bibinfo {year} {1957}{\natexlab{b}})}\BibitemShut
  {NoStop}%
\bibitem [{\citenamefont {Shannon}(1948)}]{Shannon1948}%
  \BibitemOpen
  \bibfield  {author} {\bibinfo {author} {\bibfnamefont {C.~E.}\ \bibnamefont
  {Shannon}},\ }\href@noop {} {\bibfield  {journal} {\bibinfo  {journal} {Bell
  System Tech. J.}\ }\textbf {\bibinfo {volume} {27}},\ \bibinfo {pages} {379}
  (\bibinfo {year} {1948})}\BibitemShut {NoStop}%
\bibitem [{\citenamefont {Press\'{e}}\ \emph {et~al.}(2013)\citenamefont
  {Press\'{e}}, \citenamefont {Ghosh}, \citenamefont {Lee},\ and\ \citenamefont
  {Dill}}]{Presse2013}%
  \BibitemOpen
  \bibfield  {author} {\bibinfo {author} {\bibfnamefont {S.}~\bibnamefont
  {Press\'{e}}}, \bibinfo {author} {\bibfnamefont {K.}~\bibnamefont {Ghosh}},
  \bibinfo {author} {\bibfnamefont {J.}~\bibnamefont {Lee}}, \ and\ \bibinfo
  {author} {\bibfnamefont {K.~A.}\ \bibnamefont {Dill}},\ }\href {\doibase
  10.1103/RevModPhys.85.1115} {\bibfield  {journal} {\bibinfo  {journal} {Rev.
  Mod. Phys.}\ }\textbf {\bibinfo {volume} {85}},\ \bibinfo {pages} {1115}
  (\bibinfo {year} {2013})}\BibitemShut {NoStop}%
\bibitem [{\citenamefont {Livesey}\ and\ \citenamefont
  {Brochon}(1987)}]{Livesey1987}%
  \BibitemOpen
  \bibfield  {author} {\bibinfo {author} {\bibfnamefont {A.~K.}\ \bibnamefont
  {Livesey}}\ and\ \bibinfo {author} {\bibfnamefont {J.~C.}\ \bibnamefont
  {Brochon}},\ }\href {\doibase 10.1016/S0006-3495(87)83264-2} {\bibfield
  {journal} {\bibinfo  {journal} {Biophys. J.}\ }\textbf {\bibinfo {volume}
  {52}},\ \bibinfo {pages} {693} (\bibinfo {year} {1987})}\BibitemShut
  {NoStop}%
\bibitem [{\citenamefont {Steinbach}\ \emph {et~al.}(2002)\citenamefont
  {Steinbach}, \citenamefont {Ionescu},\ and\ \citenamefont
  {Matthews}}]{Steinbach2002}%
  \BibitemOpen
  \bibfield  {author} {\bibinfo {author} {\bibfnamefont {P.~J.}\ \bibnamefont
  {Steinbach}}, \bibinfo {author} {\bibfnamefont {R.}~\bibnamefont {Ionescu}},
  \ and\ \bibinfo {author} {\bibfnamefont {C.~R.}\ \bibnamefont {Matthews}},\
  }\href {\doibase 10.1016/S0006-3495(02)75570-7} {\bibfield  {journal}
  {\bibinfo  {journal} {Biophys. J.}\ }\textbf {\bibinfo {volume} {82}},\
  \bibinfo {pages} {2244} (\bibinfo {year} {2002})}\BibitemShut {NoStop}%
\bibitem [{\citenamefont {Yeo}\ and\ \citenamefont {Burge}(2004)}]{Yeo2004}%
  \BibitemOpen
  \bibfield  {author} {\bibinfo {author} {\bibfnamefont {G.}~\bibnamefont
  {Yeo}}\ and\ \bibinfo {author} {\bibfnamefont {C.~B.}\ \bibnamefont
  {Burge}},\ }\href {\doibase 10.1089/1066527041410418} {\bibfield  {journal}
  {\bibinfo  {journal} {J. Comput. Biol.}\ }\textbf {\bibinfo {volume} {11}},\
  \bibinfo {pages} {377} (\bibinfo {year} {2004})}\BibitemShut {NoStop}%
\bibitem [{\citenamefont {Watkins}\ \emph {et~al.}(2006)\citenamefont
  {Watkins}, \citenamefont {Chang},\ and\ \citenamefont {Yang}}]{Watkins2006}%
  \BibitemOpen
  \bibfield  {author} {\bibinfo {author} {\bibfnamefont {L.~P.}\ \bibnamefont
  {Watkins}}, \bibinfo {author} {\bibfnamefont {H.}~\bibnamefont {Chang}}, \
  and\ \bibinfo {author} {\bibfnamefont {H.}~\bibnamefont {Yang}},\ }\href
  {\doibase 10.1021/jp055886d} {\bibfield  {journal} {\bibinfo  {journal} {J.
  Phys. Chem. A}\ }\textbf {\bibinfo {volume} {110}},\ \bibinfo {pages} {5191}
  (\bibinfo {year} {2006})}\BibitemShut {NoStop}%
\bibitem [{\citenamefont {Vergassola}\ \emph {et~al.}(2007)\citenamefont
  {Vergassola}, \citenamefont {Villermaux},\ and\ \citenamefont
  {Shraiman}}]{Vergassola2007}%
  \BibitemOpen
  \bibfield  {author} {\bibinfo {author} {\bibfnamefont {M.}~\bibnamefont
  {Vergassola}}, \bibinfo {author} {\bibfnamefont {E.}~\bibnamefont
  {Villermaux}}, \ and\ \bibinfo {author} {\bibfnamefont {B.~I.}\ \bibnamefont
  {Shraiman}},\ }\href {\doibase 10.1038/nature05464} {\bibfield  {journal}
  {\bibinfo  {journal} {Nature}\ }\textbf {\bibinfo {volume} {445}},\ \bibinfo
  {pages} {406} (\bibinfo {year} {2007})}\BibitemShut {NoStop}%
\bibitem [{\citenamefont {Saul}\ and\ \citenamefont {Filkov}(2007)}]{Saul2007}%
  \BibitemOpen
  \bibfield  {author} {\bibinfo {author} {\bibfnamefont {Z.~M.}\ \bibnamefont
  {Saul}}\ and\ \bibinfo {author} {\bibfnamefont {V.}~\bibnamefont {Filkov}},\
  }\href {\doibase 10.1093/bioinformatics/btm370} {\bibfield  {journal}
  {\bibinfo  {journal} {Bioinformatics}\ }\textbf {\bibinfo {volume} {23}},\
  \bibinfo {pages} {2604} (\bibinfo {year} {2007})}\BibitemShut {NoStop}%
\bibitem [{\citenamefont {Walczak}\ \emph {et~al.}(2010)\citenamefont
  {Walczak}, \citenamefont {Tka\v{c}ik},\ and\ \citenamefont
  {Bialek}}]{Walczak2010}%
  \BibitemOpen
  \bibfield  {author} {\bibinfo {author} {\bibfnamefont {A.~M.}\ \bibnamefont
  {Walczak}}, \bibinfo {author} {\bibfnamefont {G.}~\bibnamefont {Tka\v{c}ik}},
  \ and\ \bibinfo {author} {\bibfnamefont {W.}~\bibnamefont {Bialek}},\ }\href
  {\doibase 10.1103/PhysRevE.81.041905} {\bibfield  {journal} {\bibinfo
  {journal} {Phys. Rev. E}\ }\textbf {\bibinfo {volume} {81}},\ \bibinfo
  {pages} {041905} (\bibinfo {year} {2010})}\BibitemShut {NoStop}%
\bibitem [{\citenamefont {Morcos}\ \emph {et~al.}(2014)\citenamefont {Morcos},
  \citenamefont {Schafer}, \citenamefont {Cheng}, \citenamefont {Onuchic},\
  and\ \citenamefont {Wolynes}}]{Morcos2014}%
  \BibitemOpen
  \bibfield  {author} {\bibinfo {author} {\bibfnamefont {F.}~\bibnamefont
  {Morcos}}, \bibinfo {author} {\bibfnamefont {N.~P.}\ \bibnamefont {Schafer}},
  \bibinfo {author} {\bibfnamefont {R.~R.}\ \bibnamefont {Cheng}}, \bibinfo
  {author} {\bibfnamefont {J.~N.}\ \bibnamefont {Onuchic}}, \ and\ \bibinfo
  {author} {\bibfnamefont {P.~G.}\ \bibnamefont {Wolynes}},\ }\href {\doibase
  10.1073/pnas.1413575111} {\bibfield  {journal} {\bibinfo  {journal} {Proc.
  Natl. Acad. Sci. U. S. A}\ }\textbf {\bibinfo {volume} {111}},\ \bibinfo
  {pages} {12408} (\bibinfo {year} {2014})}\BibitemShut {NoStop}%
\bibitem [{\citenamefont {Shlens}\ \emph {et~al.}(2006)\citenamefont {Shlens},
  \citenamefont {Field}, \citenamefont {Gauthier}, \citenamefont {Grivich},
  \citenamefont {Petrusca}, \citenamefont {Sher}, \citenamefont {Litke},\ and\
  \citenamefont {Chichilnisky}}]{Shlens2006}%
  \BibitemOpen
  \bibfield  {author} {\bibinfo {author} {\bibfnamefont {J.}~\bibnamefont
  {Shlens}}, \bibinfo {author} {\bibfnamefont {G.~D.}\ \bibnamefont {Field}},
  \bibinfo {author} {\bibfnamefont {J.~L.}\ \bibnamefont {Gauthier}}, \bibinfo
  {author} {\bibfnamefont {M.~I.}\ \bibnamefont {Grivich}}, \bibinfo {author}
  {\bibfnamefont {D.}~\bibnamefont {Petrusca}}, \bibinfo {author}
  {\bibfnamefont {A.}~\bibnamefont {Sher}}, \bibinfo {author} {\bibfnamefont
  {A.~M.}\ \bibnamefont {Litke}}, \ and\ \bibinfo {author} {\bibfnamefont
  {E.~J.}\ \bibnamefont {Chichilnisky}},\ }\href {\doibase
  10.1523/JNEUROSCI.1282-06.2006} {\bibfield  {journal} {\bibinfo  {journal}
  {J. Neurosci.}\ }\textbf {\bibinfo {volume} {26}},\ \bibinfo {pages} {8254}
  (\bibinfo {year} {2006})}\BibitemShut {NoStop}%
\bibitem [{\citenamefont {Schneidman}\ \emph {et~al.}(2006)\citenamefont
  {Schneidman}, \citenamefont {Berry}, \citenamefont {Segev},\ and\
  \citenamefont {Bialek}}]{Schneidman2006}%
  \BibitemOpen
  \bibfield  {author} {\bibinfo {author} {\bibfnamefont {E.}~\bibnamefont
  {Schneidman}}, \bibinfo {author} {\bibfnamefont {M.~J.}\ \bibnamefont
  {Berry}}, \bibinfo {author} {\bibfnamefont {R.}~\bibnamefont {Segev}}, \ and\
  \bibinfo {author} {\bibfnamefont {W.}~\bibnamefont {Bialek}},\ }\href
  {\doibase 10.1038/nature04701} {\bibfield  {journal} {\bibinfo  {journal}
  {Nature}\ }\textbf {\bibinfo {volume} {440}},\ \bibinfo {pages} {1007}
  (\bibinfo {year} {2006})}\BibitemShut {NoStop}%
\bibitem [{\citenamefont {Tang}\ \emph {et~al.}(2008)\citenamefont {Tang},
  \citenamefont {Jackson}, \citenamefont {Hobbs}, \citenamefont {Chen},
  \citenamefont {Smith}, \citenamefont {Patel}, \citenamefont {Prieto},
  \citenamefont {Petrusca}, \citenamefont {Grivich}, \citenamefont {Sher},
  \citenamefont {Hottowy}, \citenamefont {Dabrowski}, \citenamefont {Litke},\
  and\ \citenamefont {Beggs}}]{Tang2008}%
  \BibitemOpen
  \bibfield  {author} {\bibinfo {author} {\bibfnamefont {A.}~\bibnamefont
  {Tang}}, \bibinfo {author} {\bibfnamefont {D.}~\bibnamefont {Jackson}},
  \bibinfo {author} {\bibfnamefont {J.}~\bibnamefont {Hobbs}}, \bibinfo
  {author} {\bibfnamefont {W.}~\bibnamefont {Chen}}, \bibinfo {author}
  {\bibfnamefont {J.~L.}\ \bibnamefont {Smith}}, \bibinfo {author}
  {\bibfnamefont {H.}~\bibnamefont {Patel}}, \bibinfo {author} {\bibfnamefont
  {A.}~\bibnamefont {Prieto}}, \bibinfo {author} {\bibfnamefont
  {D.}~\bibnamefont {Petrusca}}, \bibinfo {author} {\bibfnamefont {M.~I.}\
  \bibnamefont {Grivich}}, \bibinfo {author} {\bibfnamefont {A.}~\bibnamefont
  {Sher}}, \bibinfo {author} {\bibfnamefont {P.}~\bibnamefont {Hottowy}},
  \bibinfo {author} {\bibfnamefont {W.}~\bibnamefont {Dabrowski}}, \bibinfo
  {author} {\bibfnamefont {A.~M.}\ \bibnamefont {Litke}}, \ and\ \bibinfo
  {author} {\bibfnamefont {J.~M.}\ \bibnamefont {Beggs}},\ }\href {\doibase
  10.1523/JNEUROSCI.3359-07.2008} {\bibfield  {journal} {\bibinfo  {journal}
  {J. Neurosci.}\ }\textbf {\bibinfo {volume} {28}},\ \bibinfo {pages} {505}
  (\bibinfo {year} {2008})}\BibitemShut {NoStop}%
\bibitem [{\citenamefont {Marre}\ \emph {et~al.}(2009)\citenamefont {Marre},
  \citenamefont {{El Boustani}}, \citenamefont {Fr\'{e}gnac},\ and\
  \citenamefont {Destexhe}}]{Marre2009}%
  \BibitemOpen
  \bibfield  {author} {\bibinfo {author} {\bibfnamefont {O.}~\bibnamefont
  {Marre}}, \bibinfo {author} {\bibfnamefont {S.}~\bibnamefont {{El
  Boustani}}}, \bibinfo {author} {\bibfnamefont {Y.}~\bibnamefont
  {Fr\'{e}gnac}}, \ and\ \bibinfo {author} {\bibfnamefont {A.}~\bibnamefont
  {Destexhe}},\ }\href {\doibase 10.1103/PhysRevLett.102.138101} {\bibfield
  {journal} {\bibinfo  {journal} {Phys. Rev. Lett.}\ }\textbf {\bibinfo
  {volume} {102}},\ \bibinfo {pages} {138101} (\bibinfo {year}
  {2009})}\BibitemShut {NoStop}%
\bibitem [{\citenamefont {Yeh}\ \emph {et~al.}(2010)\citenamefont {Yeh},
  \citenamefont {Tang}, \citenamefont {Hobbs}, \citenamefont {Hottowy},
  \citenamefont {Dabrowski}, \citenamefont {Sher}, \citenamefont {Litke},\ and\
  \citenamefont {Beggs}}]{Yeh2010}%
  \BibitemOpen
  \bibfield  {author} {\bibinfo {author} {\bibfnamefont {F.-C.}\ \bibnamefont
  {Yeh}}, \bibinfo {author} {\bibfnamefont {A.}~\bibnamefont {Tang}}, \bibinfo
  {author} {\bibfnamefont {J.~P.}\ \bibnamefont {Hobbs}}, \bibinfo {author}
  {\bibfnamefont {P.}~\bibnamefont {Hottowy}}, \bibinfo {author} {\bibfnamefont
  {W.}~\bibnamefont {Dabrowski}}, \bibinfo {author} {\bibfnamefont
  {A.}~\bibnamefont {Sher}}, \bibinfo {author} {\bibfnamefont {A.}~\bibnamefont
  {Litke}}, \ and\ \bibinfo {author} {\bibfnamefont {J.~M.}\ \bibnamefont
  {Beggs}},\ }\href {\doibase 10.3390/e12010089} {\bibfield  {journal}
  {\bibinfo  {journal} {Entropy}\ }\textbf {\bibinfo {volume} {12}},\ \bibinfo
  {pages} {89} (\bibinfo {year} {2010})}\BibitemShut {NoStop}%
\bibitem [{\citenamefont {Stephens}\ \emph {et~al.}(2010)\citenamefont
  {Stephens}, \citenamefont {Osborne},\ and\ \citenamefont
  {Bialek}}]{Stephens2010}%
  \BibitemOpen
  \bibfield  {author} {\bibinfo {author} {\bibfnamefont {G.~J.}\ \bibnamefont
  {Stephens}}, \bibinfo {author} {\bibfnamefont {L.~C.}\ \bibnamefont
  {Osborne}}, \ and\ \bibinfo {author} {\bibfnamefont {W.}~\bibnamefont
  {Bialek}},\ }\href {\doibase 10.1073/pnas.1010868108} {\bibfield  {journal}
  {\bibinfo  {journal} {Proc. Natl. Acad. Sci. U.SA.}\ }\textbf {\bibinfo
  {volume} {108}},\ \bibinfo {pages} {15565} (\bibinfo {year}
  {2010})}\BibitemShut {NoStop}%
\bibitem [{\citenamefont {Ganmor}\ \emph {et~al.}(2011)\citenamefont {Ganmor},
  \citenamefont {Segev},\ and\ \citenamefont {Schneidman}}]{Ganmor2011}%
  \BibitemOpen
  \bibfield  {author} {\bibinfo {author} {\bibfnamefont {E.}~\bibnamefont
  {Ganmor}}, \bibinfo {author} {\bibfnamefont {R.}~\bibnamefont {Segev}}, \
  and\ \bibinfo {author} {\bibfnamefont {E.}~\bibnamefont {Schneidman}},\
  }\href {\doibase 10.1073/pnas.1019641108} {\bibfield  {journal} {\bibinfo
  {journal} {Proc. Natl. Acad. Sci. U. S. A.}\ }\textbf {\bibinfo {volume}
  {108}},\ \bibinfo {pages} {9679} (\bibinfo {year} {2011})}\BibitemShut
  {NoStop}%
\bibitem [{\citenamefont {Watanabe}\ \emph {et~al.}(2013)\citenamefont
  {Watanabe}, \citenamefont {Hirose}, \citenamefont {Wada}, \citenamefont
  {Imai}, \citenamefont {Machida}, \citenamefont {Shirouzu}, \citenamefont
  {Konishi}, \citenamefont {Miyashita},\ and\ \citenamefont
  {Masuda}}]{Watanabe2013}%
  \BibitemOpen
  \bibfield  {author} {\bibinfo {author} {\bibfnamefont {T.}~\bibnamefont
  {Watanabe}}, \bibinfo {author} {\bibfnamefont {S.}~\bibnamefont {Hirose}},
  \bibinfo {author} {\bibfnamefont {H.}~\bibnamefont {Wada}}, \bibinfo {author}
  {\bibfnamefont {Y.}~\bibnamefont {Imai}}, \bibinfo {author} {\bibfnamefont
  {T.}~\bibnamefont {Machida}}, \bibinfo {author} {\bibfnamefont
  {I.}~\bibnamefont {Shirouzu}}, \bibinfo {author} {\bibfnamefont
  {S.}~\bibnamefont {Konishi}}, \bibinfo {author} {\bibfnamefont
  {Y.}~\bibnamefont {Miyashita}}, \ and\ \bibinfo {author} {\bibfnamefont
  {N.}~\bibnamefont {Masuda}},\ }\href {\doibase 10.1038/ncomms2388} {\bibfield
   {journal} {\bibinfo  {journal} {Nat. Commun.}\ }\textbf {\bibinfo {volume}
  {4}},\ \bibinfo {pages} {1370} (\bibinfo {year} {2013})}\BibitemShut
  {NoStop}%
\bibitem [{\citenamefont {Ercsey-Ravasz}\ \emph {et~al.}(2013)\citenamefont
  {Ercsey-Ravasz}, \citenamefont {Markov}, \citenamefont {Lamy}, \citenamefont
  {{Van Essen}}, \citenamefont {Knoblauch}, \citenamefont {Toroczkai},\ and\
  \citenamefont {Kennedy}}]{Ercsey-Ravasz2013}%
  \BibitemOpen
  \bibfield  {author} {\bibinfo {author} {\bibfnamefont {M.}~\bibnamefont
  {Ercsey-Ravasz}}, \bibinfo {author} {\bibfnamefont {N.~T.}\ \bibnamefont
  {Markov}}, \bibinfo {author} {\bibfnamefont {C.}~\bibnamefont {Lamy}},
  \bibinfo {author} {\bibfnamefont {D.~C.}\ \bibnamefont {{Van Essen}}},
  \bibinfo {author} {\bibfnamefont {K.}~\bibnamefont {Knoblauch}}, \bibinfo
  {author} {\bibfnamefont {Z.}~\bibnamefont {Toroczkai}}, \ and\ \bibinfo
  {author} {\bibfnamefont {H.}~\bibnamefont {Kennedy}},\ }\href {\doibase
  10.1016/j.neuron.2013.07.036} {\bibfield  {journal} {\bibinfo  {journal}
  {Neuron}\ }\textbf {\bibinfo {volume} {80}},\ \bibinfo {pages} {184}
  (\bibinfo {year} {2013})}\BibitemShut {NoStop}%
\bibitem [{\citenamefont {Phillips}\ \emph {et~al.}(2006)\citenamefont
  {Phillips}, \citenamefont {Anderson},\ and\ \citenamefont
  {Schapire}}]{Phillips2006}%
  \BibitemOpen
  \bibfield  {author} {\bibinfo {author} {\bibfnamefont {S.~J.}\ \bibnamefont
  {Phillips}}, \bibinfo {author} {\bibfnamefont {R.~P.}\ \bibnamefont
  {Anderson}}, \ and\ \bibinfo {author} {\bibfnamefont {R.~E.}\ \bibnamefont
  {Schapire}},\ }\href {\doibase 10.1016/j.ecolmodel.2005.03.026} {\bibfield
  {journal} {\bibinfo  {journal} {Ecol. Model.}\ }\textbf {\bibinfo {volume}
  {190}},\ \bibinfo {pages} {231} (\bibinfo {year} {2006})}\BibitemShut
  {NoStop}%
\bibitem [{\citenamefont {Harte}(2011)}]{Harte2011}%
  \BibitemOpen
  \bibfield  {author} {\bibinfo {author} {\bibfnamefont {J.}~\bibnamefont
  {Harte}},\ }\href@noop {} {\emph {\bibinfo {title} {{Maximum Entropy and
  Ecology}}}}\ (\bibinfo  {publisher} {Oxford University Press},\ \bibinfo
  {year} {2011})\BibitemShut {NoStop}%
\bibitem [{\citenamefont {Fronczak}\ \emph {et~al.}(2007)\citenamefont
  {Fronczak}, \citenamefont {Fronczak},\ and\ \citenamefont
  {Ho{\l}yst}}]{Fronczak2007}%
  \BibitemOpen
  \bibfield  {author} {\bibinfo {author} {\bibfnamefont {P.}~\bibnamefont
  {Fronczak}}, \bibinfo {author} {\bibfnamefont {A.}~\bibnamefont {Fronczak}},
  \ and\ \bibinfo {author} {\bibfnamefont {J.~A.}\ \bibnamefont {Ho{\l}yst}},\
  }\href {\doibase 10.1103/PhysRevE.75.026103} {\bibfield  {journal} {\bibinfo
  {journal} {Phys. Rev. E}\ }\textbf {\bibinfo {volume} {75}},\ \bibinfo
  {pages} {026013} (\bibinfo {year} {2007})}\BibitemShut {NoStop}%
\bibitem [{\citenamefont {Wimmer}\ and\ \citenamefont
  {Lewis}(2010)}]{Wimmer2010}%
  \BibitemOpen
  \bibfield  {author} {\bibinfo {author} {\bibfnamefont {A.}~\bibnamefont
  {Wimmer}}\ and\ \bibinfo {author} {\bibfnamefont {K.}~\bibnamefont {Lewis}},\
  }\href {\doibase 10.1086/653658} {\bibfield  {journal} {\bibinfo  {journal}
  {Am. J. Sociol.}\ }\textbf {\bibinfo {volume} {116}},\ \bibinfo {pages} {583}
  (\bibinfo {year} {2010})}\BibitemShut {NoStop}%
\bibitem [{\citenamefont {Bass}(1974)}]{Bass1974}%
  \BibitemOpen
  \bibfield  {author} {\bibinfo {author} {\bibfnamefont {F.~M.}\ \bibnamefont
  {Bass}},\ }\href {\doibase 10.2307/3150989} {\bibfield  {journal} {\bibinfo
  {journal} {J. Marketing Res.}\ }\textbf {\bibinfo {volume} {11}},\ \bibinfo
  {pages} {1} (\bibinfo {year} {1974})}\BibitemShut {NoStop}%
\bibitem [{\citenamefont {Agrawal}\ \emph {et~al.}(2008)\citenamefont
  {Agrawal}, \citenamefont {Wang},\ and\ \citenamefont {Ye}}]{Agrawal2008raey}%
  \BibitemOpen
  \bibfield  {author} {\bibinfo {author} {\bibfnamefont {S.}~\bibnamefont
  {Agrawal}}, \bibinfo {author} {\bibfnamefont {Z.}~\bibnamefont {Wang}}, \
  and\ \bibinfo {author} {\bibfnamefont {Y.}~\bibnamefont {Ye}},\ }in\ \href
  {\doibase 10.1007/978-3-540-92185-1_21} {\emph {\bibinfo {booktitle}
  {Internet and Network Economics}}},\ \bibinfo {series} {Lecture Notes in
  Computer Science}, Vol.\ \bibinfo {volume} {5385},\ \bibinfo {editor} {edited
  by\ \bibinfo {editor} {\bibfnamefont {C.}~\bibnamefont {Papadimitriou}}\ and\
  \bibinfo {editor} {\bibfnamefont {S.}~\bibnamefont {Zhang}}}\ (\bibinfo
  {publisher} {Springer Berlin Heidelberg},\ \bibinfo {year} {2008})\ pp.\
  \bibinfo {pages} {126--137}\BibitemShut {NoStop}%
\bibitem [{\citenamefont {Gull}\ and\ \citenamefont
  {Daniell}(1978)}]{Gull1978}%
  \BibitemOpen
  \bibfield  {author} {\bibinfo {author} {\bibfnamefont {S.~F.}\ \bibnamefont
  {Gull}}\ and\ \bibinfo {author} {\bibfnamefont {G.~J.}\ \bibnamefont
  {Daniell}},\ }\href {\doibase 10.1038/272686a0} {\bibfield  {journal}
  {\bibinfo  {journal} {Nature}\ }\textbf {\bibinfo {volume} {272}},\ \bibinfo
  {pages} {686} (\bibinfo {year} {1978})}\BibitemShut {NoStop}%
\bibitem [{\citenamefont {Skoglund}\ \emph {et~al.}(1996)\citenamefont
  {Skoglund}, \citenamefont {Ofverstedt}, \citenamefont {Burnett},\ and\
  \citenamefont {Bricogne}}]{Skoglund1996}%
  \BibitemOpen
  \bibfield  {author} {\bibinfo {author} {\bibfnamefont {U.}~\bibnamefont
  {Skoglund}}, \bibinfo {author} {\bibfnamefont {L.~G.}\ \bibnamefont
  {Ofverstedt}}, \bibinfo {author} {\bibfnamefont {R.~M.}\ \bibnamefont
  {Burnett}}, \ and\ \bibinfo {author} {\bibfnamefont {G.}~\bibnamefont
  {Bricogne}},\ }\href {\doibase 10.1006/jsbi.1996.0081} {\bibfield  {journal}
  {\bibinfo  {journal} {J. Struct. Biol.}\ }\textbf {\bibinfo {volume} {117}},\
  \bibinfo {pages} {173} (\bibinfo {year} {1996})}\BibitemShut {NoStop}%
\bibitem [{\citenamefont {Rosenfeld}(1996)}]{Rosenfeld1996}%
  \BibitemOpen
  \bibfield  {author} {\bibinfo {author} {\bibfnamefont {R.}~\bibnamefont
  {Rosenfeld}},\ }\href {\doibase 10.1006/csla.1996.0011} {\bibfield  {journal}
  {\bibinfo  {journal} {Comput. Speech Lang.}\ }\textbf {\bibinfo {volume}
  {10}},\ \bibinfo {pages} {187} (\bibinfo {year} {1996})}\BibitemShut
  {NoStop}%
\bibitem [{\citenamefont {Holland}\ and\ \citenamefont
  {Leinhardt}(1981)}]{JAmStatAssoc_HollandL1981}%
  \BibitemOpen
  \bibfield  {author} {\bibinfo {author} {\bibfnamefont {P.~W.}\ \bibnamefont
  {Holland}}\ and\ \bibinfo {author} {\bibfnamefont {S.}~\bibnamefont
  {Leinhardt}},\ }\href@noop {} {\bibfield  {journal} {\bibinfo  {journal} {J.
  Am. Stat. Assoc.}\ }\textbf {\bibinfo {volume} {76}},\ \bibinfo {pages} {33}
  (\bibinfo {year} {1981})}\BibitemShut {NoStop}%
\bibitem [{\citenamefont {Strauss}(1986)}]{Strauss1986}%
  \BibitemOpen
  \bibfield  {author} {\bibinfo {author} {\bibfnamefont {D.}~\bibnamefont
  {Strauss}},\ }\href {\doibase 10.1137/1028156} {\bibfield  {journal}
  {\bibinfo  {journal} {SIAM Review}\ }\textbf {\bibinfo {volume} {28}},\
  \bibinfo {pages} {513} (\bibinfo {year} {1986})}\BibitemShut {NoStop}%
\bibitem [{\citenamefont {Park}\ and\ \citenamefont
  {Newman}(2004{\natexlab{a}})}]{Park2004a}%
  \BibitemOpen
  \bibfield  {author} {\bibinfo {author} {\bibfnamefont {J.}~\bibnamefont
  {Park}}\ and\ \bibinfo {author} {\bibfnamefont {M.~E.~J.}\ \bibnamefont
  {Newman}},\ }\href {\doibase 10.1103/PhysRevE.70.066146} {\bibfield
  {journal} {\bibinfo  {journal} {Phys. Rev. E}\ }\textbf {\bibinfo {volume}
  {70}},\ \bibinfo {pages} {066146} (\bibinfo {year}
  {2004}{\natexlab{a}})}\BibitemShut {NoStop}%
\bibitem [{\citenamefont {Park}\ and\ \citenamefont
  {Newman}(2004{\natexlab{b}})}]{Park2004}%
  \BibitemOpen
  \bibfield  {author} {\bibinfo {author} {\bibfnamefont {J.}~\bibnamefont
  {Park}}\ and\ \bibinfo {author} {\bibfnamefont {M.~E.~J.}\ \bibnamefont
  {Newman}},\ }\href {\doibase 10.1103/PhysRevE.70.066117} {\bibfield
  {journal} {\bibinfo  {journal} {Phys. Rev. E}\ }\textbf {\bibinfo {volume}
  {70}},\ \bibinfo {pages} {066117} (\bibinfo {year}
  {2004}{\natexlab{b}})}\BibitemShut {NoStop}%
\bibitem [{\citenamefont {Park}\ and\ \citenamefont {Newman}(2005)}]{Park2005}%
  \BibitemOpen
  \bibfield  {author} {\bibinfo {author} {\bibfnamefont {J.}~\bibnamefont
  {Park}}\ and\ \bibinfo {author} {\bibfnamefont {M.~E.~J.}\ \bibnamefont
  {Newman}},\ }\href {\doibase 10.1103/PhysRevE.72.026136} {\bibfield
  {journal} {\bibinfo  {journal} {Phys. Rev. E}\ }\textbf {\bibinfo {volume}
  {72}},\ \bibinfo {pages} {026136} (\bibinfo {year} {2005})}\BibitemShut
  {NoStop}%
\bibitem [{\citenamefont {Robins}\ \emph {et~al.}(2007)\citenamefont {Robins},
  \citenamefont {Pattison}, \citenamefont {Kalish},\ and\ \citenamefont
  {Lusher}}]{Robins2007}%
  \BibitemOpen
  \bibfield  {author} {\bibinfo {author} {\bibfnamefont {G.~L.}\ \bibnamefont
  {Robins}}, \bibinfo {author} {\bibfnamefont {P.}~\bibnamefont {Pattison}},
  \bibinfo {author} {\bibfnamefont {Y.}~\bibnamefont {Kalish}}, \ and\ \bibinfo
  {author} {\bibfnamefont {D.}~\bibnamefont {Lusher}},\ }\href {\doibase
  10.1016/j.socnet.2006.08.002} {\bibfield  {journal} {\bibinfo  {journal}
  {Soc. Networks}\ }\textbf {\bibinfo {volume} {29}},\ \bibinfo {pages} {173}
  (\bibinfo {year} {2007})}\BibitemShut {NoStop}%
\bibitem [{\citenamefont {Fronczak}\ \emph {et~al.}(2013)\citenamefont
  {Fronczak}, \citenamefont {Fronczak},\ and\ \citenamefont
  {Bujok}}]{Fronczak2013}%
  \BibitemOpen
  \bibfield  {author} {\bibinfo {author} {\bibfnamefont {P.}~\bibnamefont
  {Fronczak}}, \bibinfo {author} {\bibfnamefont {A.}~\bibnamefont {Fronczak}},
  \ and\ \bibinfo {author} {\bibfnamefont {M.}~\bibnamefont {Bujok}},\ }\href
  {\doibase 10.1103/PhysRevE.88.032810} {\bibfield  {journal} {\bibinfo
  {journal} {Phys. Rev. E}\ }\textbf {\bibinfo {volume} {88}},\ \bibinfo
  {pages} {032810} (\bibinfo {year} {2013})}\BibitemShut {NoStop}%
\bibitem [{\citenamefont {House}(2014)}]{House2014}%
  \BibitemOpen
  \bibfield  {author} {\bibinfo {author} {\bibfnamefont {T.}~\bibnamefont
  {House}},\ }\href {\doibase 10.1209/0295-5075/105/68006} {\bibfield
  {journal} {\bibinfo  {journal} {EPL}\ }\textbf {\bibinfo {volume} {105}},\
  \bibinfo {pages} {68006} (\bibinfo {year} {2014})}\BibitemShut {NoStop}%
\bibitem [{Note1()}]{Note1}%
  \BibitemOpen
  \bibinfo {note} {In the figures, however, we indicate the full range of the
  values.}\BibitemShut {Stop}%
\bibitem [{\citenamefont {Bollob\'as}(2001)}]{RndGrphs_Bollobas2001}%
  \BibitemOpen
  \bibfield  {author} {\bibinfo {author} {\bibfnamefont {B.}~\bibnamefont
  {Bollob\'as}},\ }\href@noop {} {\emph {\bibinfo {title} {Random Graphs}}},\
  \bibinfo {edition} {2nd}\ ed.\ (\bibinfo  {publisher} {Cambridge University
  Press},\ \bibinfo {year} {2001})\BibitemShut {NoStop}%
\bibitem [{\citenamefont {Kim}\ \emph {et~al.}(2009)\citenamefont {Kim},
  \citenamefont {Toroczkai}, \citenamefont {Mikl\'os}, \citenamefont
  {Erd\H{o}s},\ and\ \citenamefont {Sz\'ekely}}]{JPhysA_KimTMES2009}%
  \BibitemOpen
  \bibfield  {author} {\bibinfo {author} {\bibfnamefont {H.}~\bibnamefont
  {Kim}}, \bibinfo {author} {\bibfnamefont {Z.}~\bibnamefont {Toroczkai}},
  \bibinfo {author} {\bibfnamefont {I.}~\bibnamefont {Mikl\'os}}, \bibinfo
  {author} {\bibfnamefont {P.}~\bibnamefont {Erd\H{o}s}}, \ and\ \bibinfo
  {author} {\bibfnamefont {L.}~\bibnamefont {Sz\'ekely}},\ }\href@noop {}
  {\bibfield  {journal} {\bibinfo  {journal} {J. Phys. A}\ }\textbf {\bibinfo
  {volume} {42}},\ \bibinfo {pages} {392001} (\bibinfo {year}
  {2009})}\BibitemShut {NoStop}%
\bibitem [{\citenamefont {Del~Genio}\ \emph {et~al.}(2010)\citenamefont
  {Del~Genio}, \citenamefont {Kim}, \citenamefont {Toroczkai},\ and\
  \citenamefont {Bassler}}]{PLoSONE_DelGenioKTB2010}%
  \BibitemOpen
  \bibfield  {author} {\bibinfo {author} {\bibfnamefont {C.~I.}\ \bibnamefont
  {Del~Genio}}, \bibinfo {author} {\bibfnamefont {H.}~\bibnamefont {Kim}},
  \bibinfo {author} {\bibfnamefont {Z.}~\bibnamefont {Toroczkai}}, \ and\
  \bibinfo {author} {\bibfnamefont {K.~E.}\ \bibnamefont {Bassler}},\
  }\href@noop {} {\bibfield  {journal} {\bibinfo  {journal} {PLoS ONE}\
  }\textbf {\bibinfo {volume} {5}},\ \bibinfo {pages} {e10012} (\bibinfo {year}
  {2010})}\BibitemShut {NoStop}%
\bibitem [{\citenamefont {Czabarka}\ \emph {et~al.}(2015)\citenamefont
  {Czabarka}, \citenamefont {Dutle}, \citenamefont {Erd\H{o}s},\ and\
  \citenamefont {Mikl\'os}}]{ArXiv_CzabarkaDEM2013}%
  \BibitemOpen
  \bibfield  {author} {\bibinfo {author} {\bibfnamefont {{\'E}.}~\bibnamefont
  {Czabarka}}, \bibinfo {author} {\bibfnamefont {A.}~\bibnamefont {Dutle}},
  \bibinfo {author} {\bibfnamefont {P.~L.}\ \bibnamefont {Erd\H{o}s}}, \ and\
  \bibinfo {author} {\bibfnamefont {I.}~\bibnamefont {Mikl\'os}},\ }\href
  {\doibase http://dx.doi.org/10.1016/j.dam.2014.10.012} {\bibfield  {journal}
  {\bibinfo  {journal} {Discrete Appl. Math.}\ }\textbf {\bibinfo {volume}
  {181}},\ \bibinfo {pages} {283 } (\bibinfo {year} {2015})}\BibitemShut
  {NoStop}%
\bibitem [{\citenamefont {Jerrum}\ \emph {et~al.}(1986)\citenamefont {Jerrum},
  \citenamefont {Valiant},\ and\ \citenamefont
  {Vazirani}}]{TheorCompSci_JerrumVV1986}%
  \BibitemOpen
  \bibfield  {author} {\bibinfo {author} {\bibfnamefont {M.~R.}\ \bibnamefont
  {Jerrum}}, \bibinfo {author} {\bibfnamefont {L.~G.}\ \bibnamefont {Valiant}},
  \ and\ \bibinfo {author} {\bibfnamefont {V.~V.}\ \bibnamefont {Vazirani}},\
  }\href@noop {} {\bibfield  {journal} {\bibinfo  {journal} {Theoret. Comput.
  Sci.}\ }\textbf {\bibinfo {volume} {43}},\ \bibinfo {pages} {169} (\bibinfo
  {year} {1986})}\BibitemShut {NoStop}%
\bibitem [{\citenamefont {Vazirani}(2003)}]{ApproxAlg2003}%
  \BibitemOpen
  \bibfield  {author} {\bibinfo {author} {\bibfnamefont {V.}~\bibnamefont
  {Vazirani}},\ }\href@noop {} {\emph {\bibinfo {title} {Approximation
  Algorithms}}}\ (\bibinfo  {publisher} {Springer},\ \bibinfo {year}
  {2003})\BibitemShut {NoStop}%
\bibitem [{Note2()}]{Note2}%
  \BibitemOpen
  \bibinfo {note} {See Supplemental Material at [URL will be inserted by
  publisher], which includes Refs.~\cite
  {Ramond2010,Bona2004,Agrawal2008raey,Gleiser2003}}\BibitemShut {NoStop}%
\bibitem [{\citenamefont {Ramond}(2010)}]{Ramond2010}%
  \BibitemOpen
  \bibfield  {author} {\bibinfo {author} {\bibfnamefont {P.}~\bibnamefont
  {Ramond}},\ }\href@noop {} {\emph {\bibinfo {title} {Group Theory: A
  Physicist's Survey}}}\ (\bibinfo  {publisher} {Cambridge},\ \bibinfo {year}
  {2010})\BibitemShut {NoStop}%
\bibitem [{\citenamefont {B\'ona}(2004)}]{Bona2004}%
  \BibitemOpen
  \bibfield  {author} {\bibinfo {author} {\bibfnamefont {M.}~\bibnamefont
  {B\'ona}},\ }\href@noop {} {\emph {\bibinfo {title} {Combinatorics of
  Permutations (Discrete Mathematics and its Applications)}}}\ (\bibinfo
  {publisher} {Chapman \& Hall/CRC Press},\ \bibinfo {year} {2004})\BibitemShut
  {NoStop}%
\bibitem [{\citenamefont {Gleiser}\ and\ \citenamefont
  {Danon}(2003)}]{Gleiser2003}%
  \BibitemOpen
  \bibfield  {author} {\bibinfo {author} {\bibfnamefont {P.~M.}\ \bibnamefont
  {Gleiser}}\ and\ \bibinfo {author} {\bibfnamefont {L.}~\bibnamefont
  {Danon}},\ }\href {\doibase 10.1142/S0219525903001067} {\bibfield  {journal}
  {\bibinfo  {journal} {Adv. Complex Syst.}\ }\textbf {\bibinfo {volume} {6}},\
  \bibinfo {pages} {565} (\bibinfo {year} {2003})}\BibitemShut {NoStop}%
\bibitem [{\citenamefont {Boyd}\ and\ \citenamefont
  {Vandenberghe}(2004)}]{ConvexOpt_BoydV2004}%
  \BibitemOpen
  \bibfield  {author} {\bibinfo {author} {\bibfnamefont {S.}~\bibnamefont
  {Boyd}}\ and\ \bibinfo {author} {\bibfnamefont {L.}~\bibnamefont
  {Vandenberghe}},\ }\href@noop {} {\emph {\bibinfo {title} {Convex
  Optimization}}}\ (\bibinfo  {publisher} {Cambridge University Press},\
  \bibinfo {year} {2004})\BibitemShut {NoStop}%
\bibitem [{Note3()}]{Note3}%
  \BibitemOpen
  \bibinfo {note} {Equivalent to the usual definition for strict concavity
  $g(\lambda \protect \mathbf {x}_1 + (1-\lambda ) \protect \mathbf {x}_2) < t
  g(\protect \mathbf {x}_1) + (1-t)g(\protect \mathbf {x}_1)$, $\forall
  \protect \mathbf {x}_i \in {\protect \cal D}, \lambda \in (0,1)$ with $g =
  \protect \qopname \relax o{ln}{\protect \cal N}$}\BibitemShut {NoStop}%
\bibitem [{\citenamefont {Snijders}\ \emph {et~al.}(2006)\citenamefont
  {Snijders}, \citenamefont {Pattison}, \citenamefont {Robins},\ and\
  \citenamefont {Handcock}}]{Snijders2006}%
  \BibitemOpen
  \bibfield  {author} {\bibinfo {author} {\bibfnamefont {T.~A.~B.}\
  \bibnamefont {Snijders}}, \bibinfo {author} {\bibfnamefont {P.~E.}\
  \bibnamefont {Pattison}}, \bibinfo {author} {\bibfnamefont {G.~L.}\
  \bibnamefont {Robins}}, \ and\ \bibinfo {author} {\bibfnamefont {M.~S.}\
  \bibnamefont {Handcock}},\ }\href {\doibase 10.1111/j.1467-9531.2006.00176.x}
  {\bibfield  {journal} {\bibinfo  {journal} {Sociol. Methodol.}\ }\textbf
  {\bibinfo {volume} {36}},\ \bibinfo {pages} {99} (\bibinfo {year}
  {2006})}\BibitemShut {NoStop}%
\bibitem [{Note4()}]{Note4}%
  \BibitemOpen
  \bibinfo {note} {Since $\protect \mathbf {F}$ is bijective, the counts of
  states with $\protect \bm {\xi }$ is the same as the the counts of states
  with $\protect \ensuremath {\protect \mathbf {m}}=\protect \mathbf
  {F^{-1}}(\protect \bm {\xi })$.}\BibitemShut {Stop}%
\bibitem [{\citenamefont {Zachary}(1977)}]{Zachary1977}%
  \BibitemOpen
  \bibfield  {author} {\bibinfo {author} {\bibfnamefont {W.}~\bibnamefont
  {Zachary}},\ }\href {http://www.maths.tcd.ie/~mnl/store/Zachary1977a.pdf}
  {\bibfield  {journal} {\bibinfo  {journal} {J. Anthropol. Res.}\ }\textbf
  {\bibinfo {volume} {33}},\ \bibinfo {pages} {452} (\bibinfo {year}
  {1977})}\BibitemShut {NoStop}%
\end{thebibliography}%

\end{document}